\documentclass[bibyear]{aa} 
\usepackage{graphicx}
\usepackage{amsmath}
\usepackage{amsfonts}
\usepackage{tikz}
\usetikzlibrary{arrows.meta,fit,positioning,shapes.geometric}
\usepackage{subcaption}
\usepackage{natbib}
\bibpunct{(}{)}{;}{a}{}{,} 
\usepackage{hyperref}
\hypersetup{
  colorlinks = true, 
  urlcolor  = blue, 
  linkcolor = red,  
  citecolor = blue  
}
\usepackage{placeins}
\usepackage[varg]{txfonts}
\usepackage{orcidlink}

\title{Inferring Unreported Measurement Uncertainties via Information Geometry in Astrophysics}
\titlerunning{Inferring Unreported Measurement Uncertainties}
\author{Marko Imbri\v{s}ak\thanks{\emph{marko.imbrisak@gmail.com}}\inst{1}\orcidlink{0000-0002-2773-8617}
\and Kre\v{s}imir Tisani\'c\inst{1}\orcidlink{0000-0001-6382-4937}}

\institute{Independent researcher, Zagreb, Croatia}

\abstract{Modern radio and multi-instrument astrophysical datasets are increasingly assembled from surveys with different sensitivities, angular resolutions, frequency coverages, and selection effects. In such heterogeneous datasets, published measurement uncertainties are often incomplete, non-uniform across subsets, or missing cross-correlation information altogether. This limits reliable statistical inference, since underestimated or inconsistently modeled uncertainties can distort fitted spectral shapes, bias parameter estimates, and obscure physically meaningful structure.

We introduce the Fisher Information Metric Error Reconstruction (FIMER), an information-geometric framework for reconstructing effective measurement uncertainties directly from heterogeneous astrophysical data. FIMER combines weighted Fisher-information geometry, the Full Bayesian Evaluation Technique, and an adaptive discrete hyperparameter search, while incorporating prior statistical knowledge of detector behavior into the weighting procedure. The priors used are not chosen as arbitrary tuning prescriptions or uninformative regularizers; rather, they are motivated by plausible statistical properties of the underlying detection process. In particular, Poisson priors represent counting-statistics behavior, while extreme-value priors allow tail-dominated fluctuations to be incorporated when rare or asymmetric excursions are expected to influence the inferred uncertainty distribution. The resulting framework is interpretable, in the sense that prior statistical knowledge enters the reconstruction in an explicit and testable form.

We apply FIMER to radio spectral energy distributions of radio-excess active galactic nuclei using COSMOS VLA data at 1.4 and 3 GHz together with GMRT data at 325 and 610 MHz. The results show that FIMER provides a practical route to uncertainty reconstruction in heterogeneous survey combinations, especially when reported uncertainties are unavailable, underestimated, or strongly correlated. The method is particularly relevant for archival and multi-survey astrophysical datasets, where full covariance information is rarely available but reliable statistical inference remains essential.}

\keywords{methods: data analysis -- methods: statistical -- galaxies: active -- radio continuum: galaxies}

\begin{document}
\maketitle
\nolinenumbers
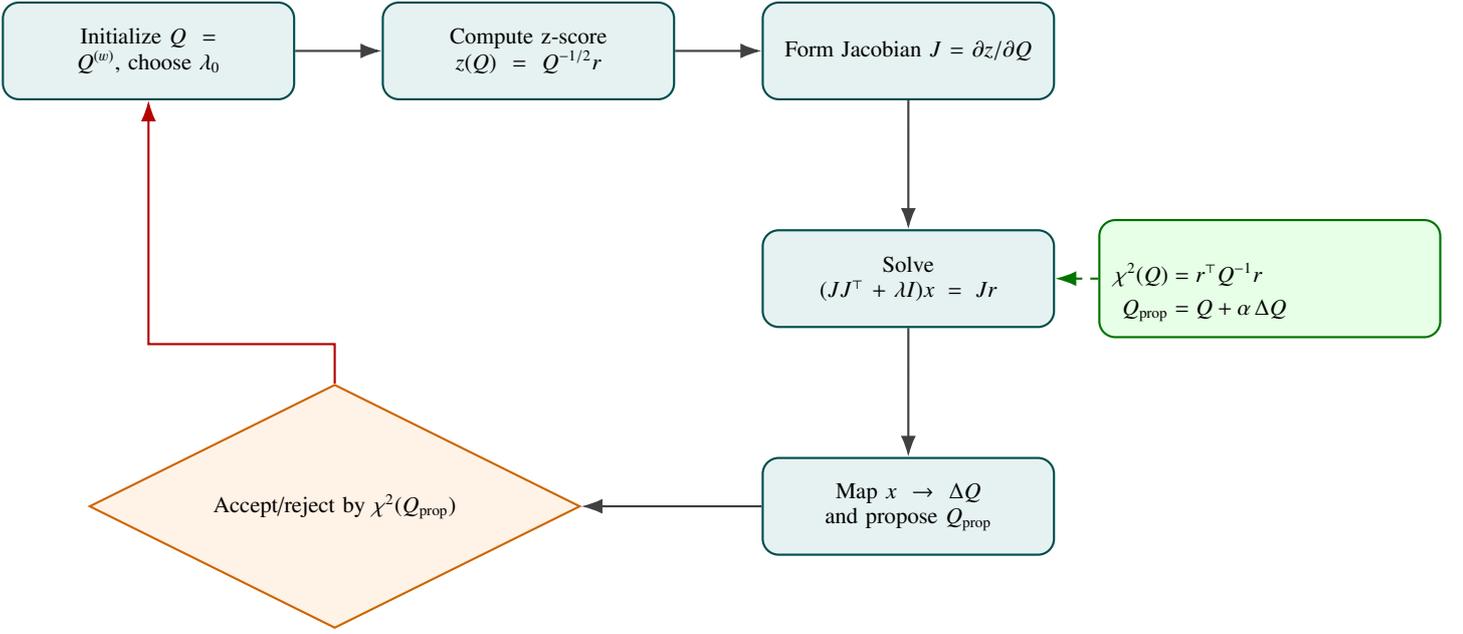
\begin{figure*}[h]
\centering
\begin{tikzpicture}[
    scale=0.95,
    transform shape,
    >=Latex,
    every node/.style={font=\small},
    proc/.style={
      rectangle,
      rounded corners=6pt,
      draw=blue!65!black,
      thick,
      fill=blue!8,
      text width=3.8cm,
      minimum height=1.35cm,
      align=center
    },
    calc/.style={
      rectangle,
      rounded corners=6pt,
      draw=teal!60!black,
      thick,
      fill=teal!10,
      text width=3.8cm,
      minimum height=1.35cm,
      align=center
    },
    decision/.style={
      diamond,
      aspect=2.0,
      draw=orange!80!black,
      thick,
      fill=orange!10,
      text width=3.4cm,
      align=center,
      inner xsep=0.25cm,
      inner ysep=0.55cm
    },
    note/.style={
      rectangle,
      rounded corners=6pt,
      draw=green!45!black,
      thick,
      fill=green!10,
      text width=4.3cm,
      align=center,
      inner sep=6pt
    },
    flow/.style={-{Latex[length=2.8mm,width=2.0mm]}, thick, draw=black!75},
    support/.style={-{Latex[length=2.8mm,width=2.0mm]}, thick, dashed, draw=green!45!black},
    feedback/.style={-{Latex[length=2.8mm,width=2.0mm]}, thick, draw=red!70!black}
  ]
    \node[calc] (a) at (0,0) {Initialize \(Q=Q^{(w)}\), choose \(\lambda_0\)};
    \node[calc,right=1.2cm of a] (b) {Compute z-score \(z(Q)=Q^{-1/2}r\)};
    \node[calc,right=1.2cm of b] (c) {Form Jacobian \(J=\partial z/\partial Q\)};
    \node[calc,below=1.8cm of c] (d) {Solve\\ \((JJ^\top+\lambda I)x=Jr\)};
    \node[calc,below=1.8cm of d] (e) {Map \(x \to \Delta Q\) and propose \(Q_{\mathrm{prop}}\)};
    \node[decision,left=2.5cm of e] (f) {Accept/reject by \(\chi^2(Q_{\mathrm{prop}})\)};
    \node[note,right=0.6cm of d] (g) {%
    \begin{align*}
        \chi^2(Q)&=r^\top Q^{-1}r\\
        Q_{\mathrm{prop}}&=Q+\alpha\,\Delta Q
    \end{align*}
    };

    \draw[flow] (a.east) -- (b.west);
    \draw[flow] (b.east) -- (c.west);
    \draw[flow] (c.south) -- (d.north);
    \draw[flow] (d.south) -- (e.north);
    \draw[flow] (e.west) -- (f.east);
    \draw[support] (g.west) -- (d.east);
    \draw[feedback] (f.north) -- ++(0,0.55) -| ([yshift=-0.28cm]a.south) -- (a.south);
\end{tikzpicture}
\caption{Schematic overview of the covariance minimization loop. The algorithm iteratively proposes symmetric updates \(\Delta Q\), evaluates objective function, and adapts the damping parameter until convergence.}\label{fig:cov_min_schema}
\end{figure*}

\begin{figure*}[h]
\centering
\begin{tikzpicture}[
    scale=0.86,
    transform shape,
    >=Latex,
    every node/.style={font=\small},
    proc/.style={
      rectangle,
      rounded corners=6pt,
      draw=blue!65!black,
      thick,
      fill=blue!8,
      text width=2.8cm,
      minimum height=1.6cm,
      align=center
    },
    calc/.style={
      rectangle,
      rounded corners=6pt,
      draw=teal!60!black,
      thick,
      fill=teal!10,
      text width=2.8cm,
      minimum height=1.6cm,
      align=center
    },
    note/.style={
      rectangle,
      rounded corners=6pt,
      draw=green!45!black,
      thick,
      fill=green!10,
      align=center,
      inner sep=6pt
    },
    flow/.style={-{Latex[length=2.8mm,width=2.0mm]}, thick, draw=black!75},
    support/.style={-{Latex[length=2.8mm,width=2.0mm]}, thick, dashed, draw=green!45!black}
  ]
    \node[proc] (a) at (0,0) {Build interpolation matrix \(S\) and set initial \(B\)};
    \node[calc,right=0.55cm of a] (b) {Compute wFIM weights \(w_k(\theta)\) };
    \node[proc,right=0.55cm of b] (c) {Form \(B^{(w)}\) and \(Q^{(w)}\)};
    \node[calc,right=0.55cm of c] (d) {Minimize $\chi^2(Q^{(w)})$};
    \node[calc,right=0.55cm of d] (e) {Pushforward \(S^\top A_1 S\)};
    \node[calc,right=0.55cm of e] (f) { Estimate measurement uncertainties};

    \node[note,text width=2.8cm,above=1.7cm of c] (g) {%
    \begin{align*}
Q^{(w)}&=S A_0 S^\top + B^{(w)}\\
B^{(w)}_{ij}&=\frac{\sigma_i^2}{w_{k(i)}(\theta)}\,\delta_{ij}
    \end{align*}
    };
    \node[proc,text width=6.2cm,below=1.7cm of e] (h) {%
    \begin{align*}
        y_1&=y_0 + A_0 S^\top(SA_0S^\top + Q)^{-1}(y_m - Sy_0)\\
        A_1 &=A_0 - A_0 S^\top(SA_0S^\top + Q)^{-1}S A_0
    \end{align*}
    };
    \node[note,text width=2.8cm,below=1.7cm of b] (i) {Choose prior distribution $f(\alpha)$
    };

    \draw[flow] (a.east) -- (b.west);
    \draw[flow] (b.east) -- (c.west);
    \draw[flow] (c.east) -- (d.west);
    \draw[flow] (d.east) -- (e.west);
    \draw[flow] (e.east) -- (f.west);
    \draw[support] (g.south) -- (c.north);
    \draw[support] (h.north) -- (e.south);
    \draw[support] (i.north) -- (b.south);
\end{tikzpicture}
\caption{Schematic overview of the \textsc{FIMER} procedure. Blue callouts indicate the core FBET steps, while green callouts highlight the weighted-covariance construction and the FBET update.}\label{fig:fimbet_schema}
\end{figure*}
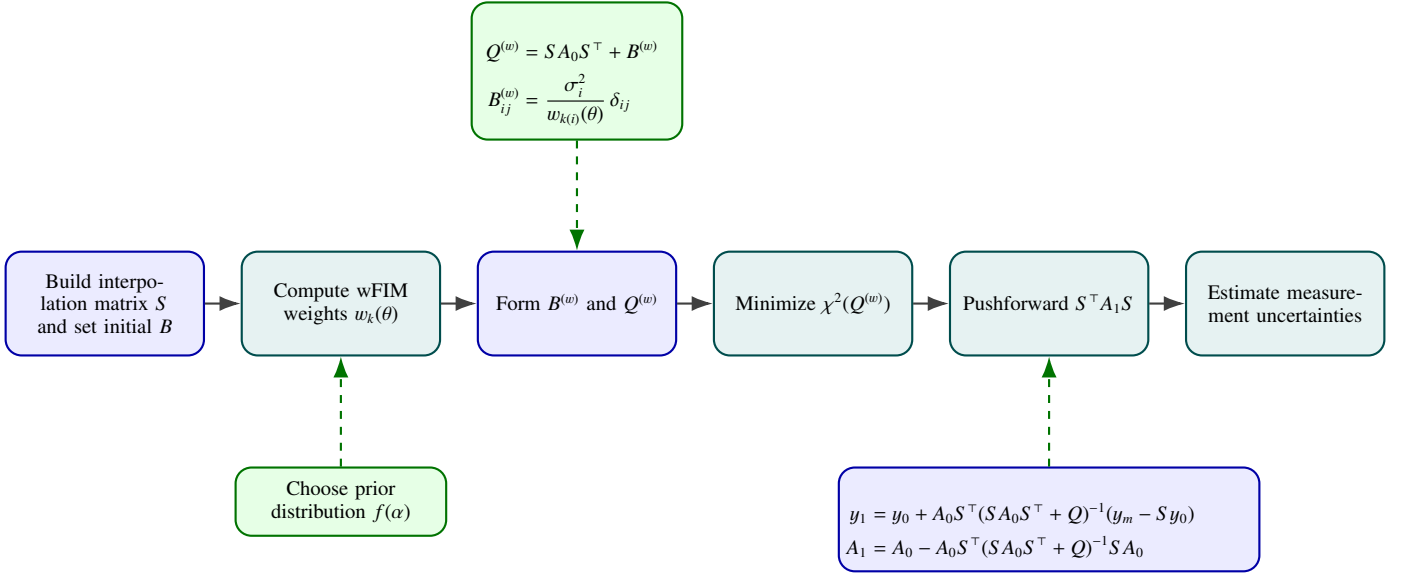
\begin{figure*}[t]
    \centering
    \resizebox{0.96\textwidth}{!}{%
    \begin{tikzpicture}[
        x=0.46cm,
        y=0.46cm,
        line cap=round,
        line join=round,
        every node/.style={font=\small},
        border/.style={draw=black!60, dashed, line width=0.9pt, rounded corners=5pt},
        coarseLine/.style={draw=blue!70!black, line width=2.2pt},
        fineLine/.style={draw=red!75!black, line width=1.8pt},
        coarsePt/.style={circle, fill=blue!75!black, draw=white, line width=0.45pt, minimum size=6.2pt, inner sep=0pt},
        finePt/.style={circle, fill=red!75!black, draw=white, line width=0.45pt, minimum size=5.2pt, inner sep=0pt},
        coarseLbl/.style={rounded corners=3pt, fill=white, draw=blue!35!black, inner sep=2pt},
        fineLbl/.style={rounded corners=3pt, fill=white, draw=red!35!black, inner sep=2pt},
        localHalo/.style={draw=red!35!black, fill=red!6, rounded corners=4pt, line width=0.9pt}
    ]
        \def\boxHalfx{12}
        \def\boxHalfy{5}
        \def\smallCx{6}
        \def\smallCy{2}
        \def\smallDx{1.5}
        \def\smallDy{1.5}

        \coordinate (O) at (0,0);
        \coordinate (L) at (-\boxHalfx,0);
        \coordinate (R) at (\boxHalfx,0);
        \coordinate (T) at (0,\boxHalfy);
        \coordinate (B) at (0,-\boxHalfy);

        \coordinate (Csmall) at (\smallCx,\smallCy);
        \coordinate (Lsmall) at (\smallCx-\smallDx,\smallCy);
        \coordinate (Rsmall) at (\smallCx+\smallDx,\smallCy);
        \coordinate (Tsmall) at (\smallCx,\smallCy+\smallDy);
        \coordinate (Bsmall) at (\smallCx,\smallCy-\smallDy);

        \path[fill=black!2, draw=none, rounded corners=5pt] (-\boxHalfx-0.7,-\boxHalfy-0.7) rectangle (\boxHalfx+0.7,\boxHalfy+0.7);
        \draw[border] (-\boxHalfx,-\boxHalfy) rectangle (\boxHalfx,\boxHalfy);
        \draw[localHalo] (\smallCx-\smallDx-0.7,\smallCy-\smallDy-0.7) rectangle (\smallCx+\smallDx+0.7,\smallCy+\smallDy+0.7);

        \draw[coarseLine] (L) -- (R);
        \draw[coarseLine] (B) -- (T);
        \draw[fineLine] (Lsmall) -- (Rsmall);
        \draw[fineLine] (Bsmall) -- (Tsmall);

        \node[coarsePt] (Odot) at (O) {};
        \node[coarsePt] (Ldot) at (L) {};
        \node[coarsePt] (Rdot) at (R) {};
        \node[coarsePt] (Tdot) at (T) {};
        \node[coarsePt] (Bdot) at (B) {};

        \node[finePt] (Cdot) at (Csmall) {};
        \node[finePt] (Lsdot) at (Lsmall) {};
        \node[finePt] (Rsdot) at (Rsmall) {};
        \node[finePt] (Tsdot) at (Tsmall) {};
        \node[finePt] (Bsdot) at (Bsmall) {};

        \node[coarseLbl,above=6pt of Tdot] {$\left(h_1^{(c)},h_2^{(\max)}\right)$};
        \node[coarseLbl,below=6pt of Bdot] {$\left(h_1^{(c)},h_2^{(\min)}\right)$};
        \node[coarseLbl,left=6pt of Ldot] {$\left(h_1^{(\min)},h_2^{(c)}\right)$};
        \node[coarseLbl,right=6pt of Rdot] {$\left(h_1^{(\max)},h_2^{(c)}\right)$};
        \node[coarseLbl,below right=2pt and 4pt of Odot] {$\left(h_1^{(c)},h_2^{(c)}\right)$};

        \node[fineLbl,above right=2pt and 4pt of Cdot] {$\left(h_1^{(t)},h_2^{(t)}\right)$};
        \node[fineLbl,left=6pt of Lsdot] {$\left(h_1^{(t)}-1,h_2^{(t)}\right)$};
        \node[fineLbl,right=6pt of Rsdot] {$\left(h_1^{(t)}+1,h_2^{(t)}\right)$};
        \node[fineLbl,above=6pt of Tsdot] {$\left(h_1^{(t)},h_2^{(t)}+1\right)$};
        \node[fineLbl,below=6pt of Bsdot] {$\left(h_1^{(t)},h_2^{(t)}-1\right)$};
    \end{tikzpicture}%
    }
    \caption{Schematic of the adaptive neighborhood search. The larger cross (blue) marks the coarse search around the central point, while the smaller cross (red) illustrates a refinement after $t$ steps.}
    \label{fig:adaptive_cross_sketch}
\end{figure*}
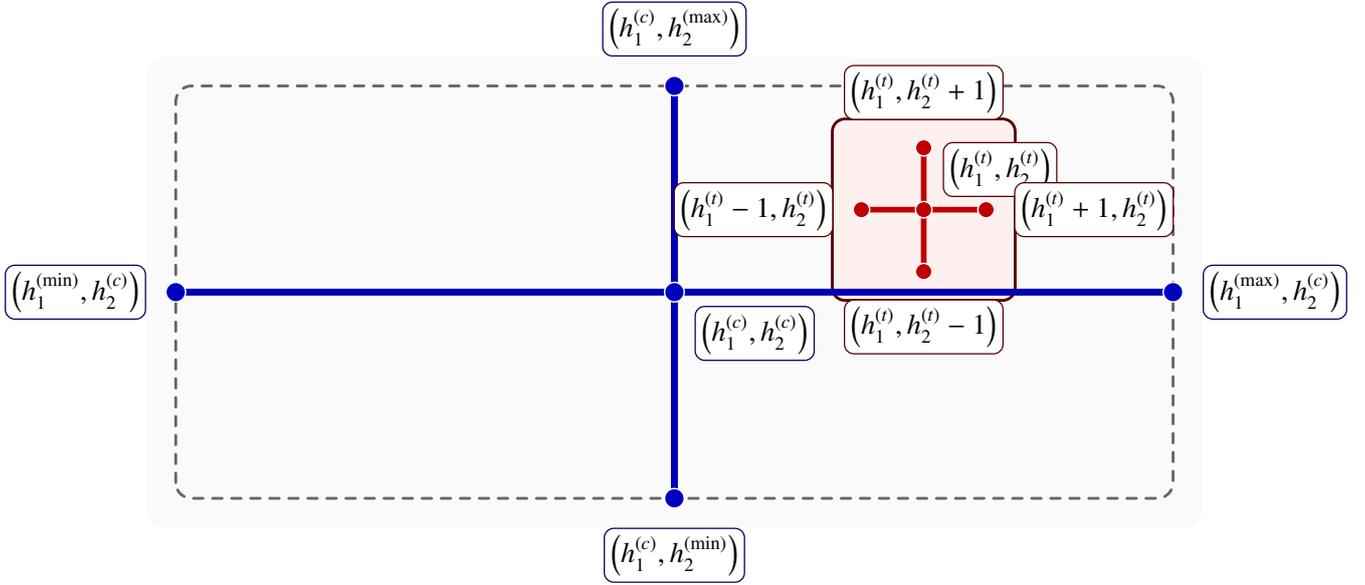

\begin{figure*}[t]
    \centering
    \begin{subfigure}[t]{\textwidth}
        \centering
        \includegraphics[width=\linewidth]{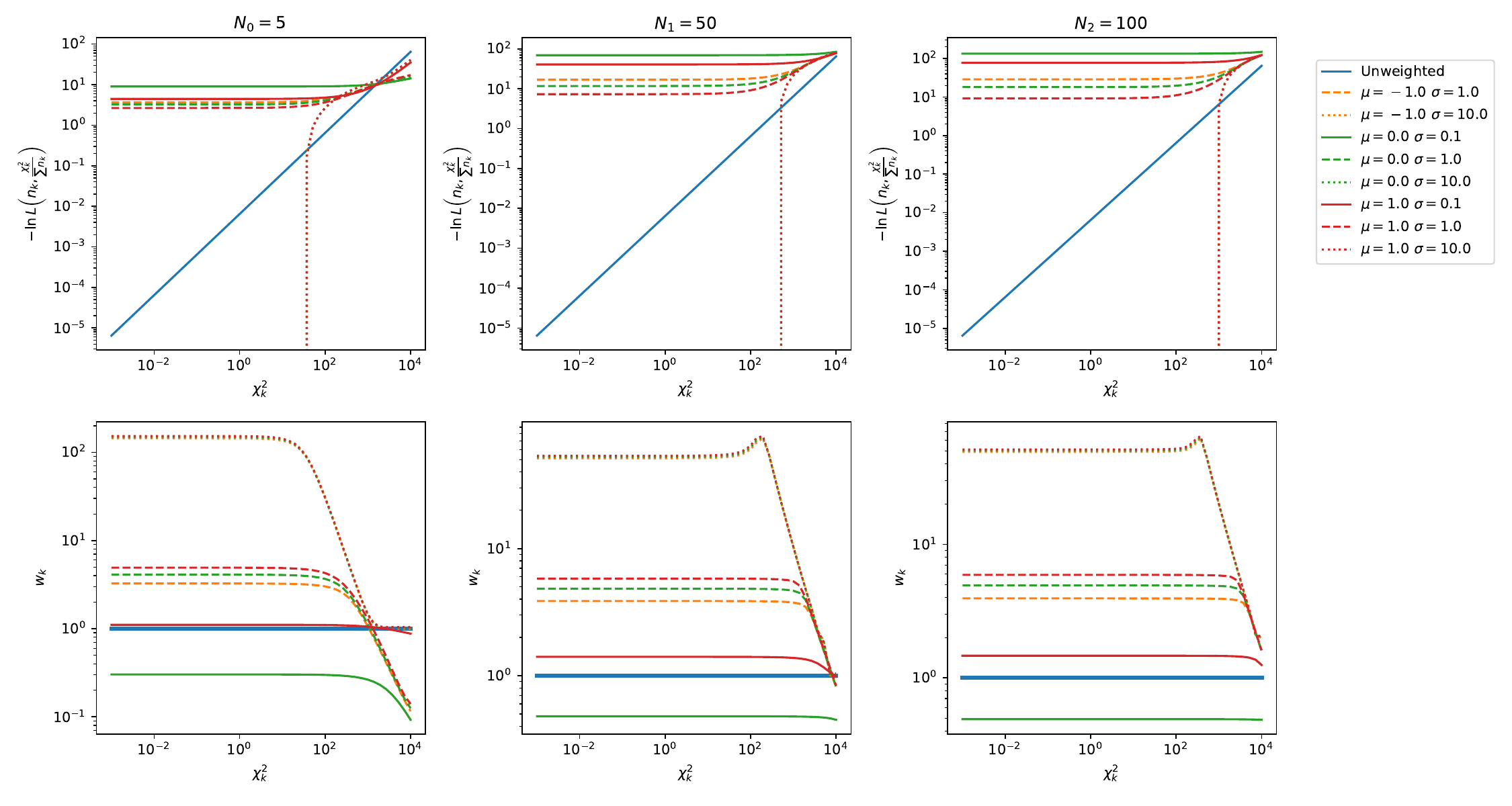}
        \caption{Extreme-value-distribution prior. The first row shows the weighted negative log-likelihoods for groups of sizes 5, 50, and 100, while the second row shows the corresponding weighted Fisher-information-metric weights.}
    \end{subfigure}

    \vspace{0.75em}

    \begin{subfigure}[t]{\textwidth}
        \centering
        \includegraphics[width=\linewidth]{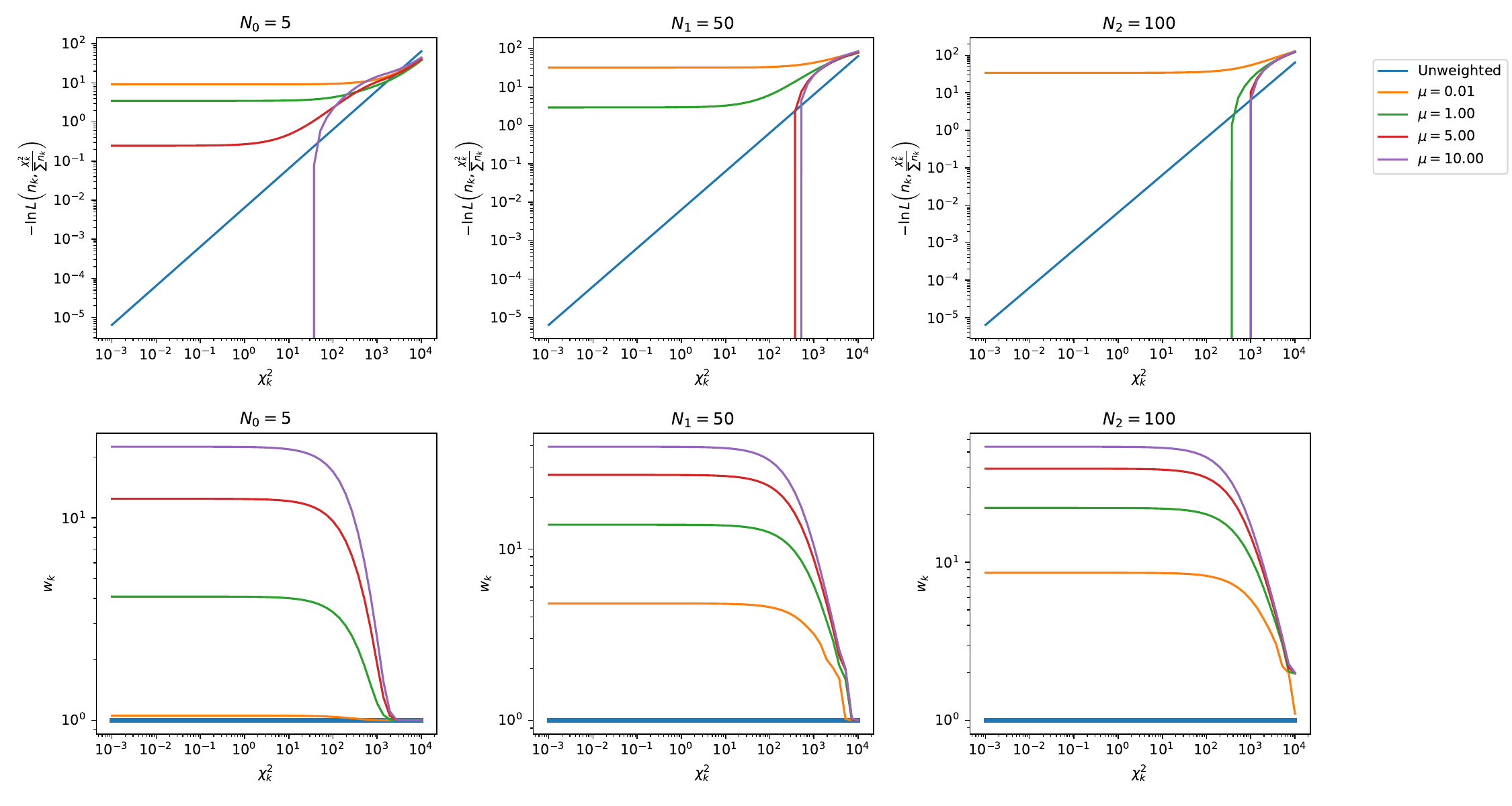}
        \caption{Poisson prior. The first row shows the weighted negative log-likelihoods for groups of sizes 5, 50, and 100, while the second row shows the corresponding weighted Fisher-information-metric weights.}
    \end{subfigure}
    \caption{Comparison of prior-dependent likelihood and metric weighting behavior for three groups of different sizes. In both subfigures, the blue curve denotes the unweighted case and the colored curves denote different prior-parameter choices.}
    \label{fig:prior_weights}
    
\end{figure*}

\section{Introduction}

{Modern astrophysical analyses increasingly rely on combining measurements from multiple surveys, instruments, and observing strategies within a single statistical study.} In the simplest case, heterogeneous datasets can be fitted by assigning different measurement variances to different groups in the \(\chi^2\) objective function. \citet{Hobson2002} introduced a method for joint analyses of cosmological datasets in which each dataset is weighted according to its own statistical properties. In many public datasets, however, uncertainty information is incomplete, most notably with respect to cross-correlations between datasets, and this omission can systematically skew parameter estimates. The Full Bayesian Evaluation Technique (FBET) \citep{leeb_consistent_2008,leeb_geneus_2011} mitigates this limitation through a generalized least-squares update that reconstructs a full covariance estimate on a chosen grid while consistently incorporating both experimental and model-driven uncertainty contributions \citep{neudecker_advanced_2014,neudecker_impact_2013}. A key challenge in this setting is that detector channels may be similarly distributed at the marginal level while remaining \emph{mutually dependent}. This makes the {prior uncertainty description} nontrivial and motivates explicit correlation modeling together with adaptive hyperparameter inference.

A recent application of information geometry is the weighted Fisher geometry developed in the weighted Levenberg--Marquardt (wLM) framework for heterogeneous multichannel fitting problems \citep{Imbrisak2025wLM}. The central idea is to replace the standard Fisher Information Metric (FIM) by a \emph{weighted} Fisher Information Metric (wFIM), obtained by marginalizing over weighting priors. This construction yields an adaptive metric on parameter space that balances the influence of heterogeneous data groups and improves parameter estimation in sloppy nonlinear models \citep{Imbrisak2025wLM}. 

Radio continuum surveys now probe galaxy and active galactic nucleus (AGN) populations over wide sky areas and broad frequency ranges, making radio spectral energy distributions (SEDs) a central tool for interpreting source physics and population evolution. In many practical analyses, radio SEDs are approximated by a single power law with a fixed spectral index, often near \(\alpha\approx0.7\) \citep{Condon92}. However, observational studies have repeatedly shown that real sources exhibit departures from this simplified picture, including curvature, spectral breaks, flattening, and turnover behavior across frequency bands and source classes \citep{Condon91,Kukula98, Kimball08, Murphy06, CalistroRivera17,Tisanic19,Tisanic2020TheNuclei}. As a result, inference based on rigid spectral assumptions can hide physically meaningful structure in heterogeneous survey data.

This issue is particularly relevant in modern multi-instrument datasets that combine measurements with different sensitivities, resolutions, frequency coverages, and detection limits. Measurement subsets at some observer-frame frequencies strongly constrain the fit while others contribute weakly; non-detections and censored points are common; and {reported uncertainties can differ substantially across subsets or be unavailable in a form suitable for a joint statistical treatment.} 
{These features make uncertainty specification and covariance propagation central to radio-SED modeling and, more broadly, to the statistical analysis of heterogeneous datasets.}

Motivated by the widespread presence of heterogeneous datasets and the advent of multi-messenger astrophysics \citep{Meszaros2019}, we introduce the Fisher Information Metric Error Reconstruction (\textsc{FIMER}), an information-geometric framework based on FBET for {reconstructing effective measurement uncertainties} in heterogeneous data. {A central point of the method is that the weighting priors are not introduced as arbitrary regularization choices or as purely empirical tuning parameters. Rather, they encode prior statistical knowledge about detector behavior and thereby allow assumptions about the underlying measurement process to enter the reconstruction explicitly. In particular, Poisson-type weighting is motivated by counting-statistics behavior expected for detector responses, while extreme-value priors provide a way to represent regimes in which tail-dominated effects, rare excursions, or asymmetric fluctuations may influence the inferred distribution of measurement uncertainties. In this way, \textsc{FIMER} is a statistically interpretable procedure in which prior knowledge of detector statistics is incorporated directly into the reconstruction.}

\textsc{FIMER} {is especially useful when reported uncertainties are unavailable, suspected to be underestimated, or strongly correlated. It is intended not as a replacement for detailed instrument-level calibration, but as a statistically coherent procedure for situations in which complete uncertainty information is unavailable or difficult to propagate consistently through the analysis.}

To guide the reader, Section~\ref{sec:IG} develops the information-geometric framework of the method, including the notation, weighted Fisher geometry, the relation to FBET, covariance minimization, the \textsc{FIMER} workflow, and the adaptive grid search. Section~\ref{sec:data} then describes the RxAGN dataset used in the application. Section~\ref{sec:results} presents the main astrophysical results, with emphasis on uncertainty reconstruction, residual diagnostics, and correlation structure. Finally, Section~\ref{sec:conclusion} summarizes the main conclusions and outlines directions for future work.

\section{Information Geometry}\label{sec:IG}
In Section~\ref{sec:notation}, we introduce the notation and the basic information-geometric setup. Sections~\ref{sec:wFIM} and~\ref{sec:FBET} then summarize the two main components combined in our framework, namely wFIM weighting and FBET. Section~\ref{sec:cov_min} presents the covariance-minimization algorithm, Section~\ref{sec:fimer} introduces the \textsc{FIMER} workflow, and Section~\ref{sec:ags} describes the hyperparameter-search procedure used to determine the optimal prior-distribution hyperparameters.
\subsection{Notation and basic information-geometric setup}\label{sec:notation}

Let \(\theta=(\theta^1,\dots,\theta^{N_p})\) denote the parameter vector of a nonlinear model. We divide the dataset into \(N_g\) groups indexed by \(k \in \{1,\dots,N_g\}\). In the present application, a group may represent an object's normalized flux densities for observations taken at different observer-frame frequencies or any other structured subset whose influence should be controlled during fitting. We denote the \(i\)-th observation in group \(k\) by \(y_i^{(k)}\), the corresponding model prediction by \(f_i^{(k)}(\theta)\), and the associated standard uncertainty by \(\sigma_i^{(k)}\). Following \citep{Imbrisak2025wLM}, the normalized residual is
\begin{equation}
r_i^{(k)}(\theta)=
\frac{y_i^{(k)}-f_i^{(k)}(\theta)}{\sigma_i^{(k)}}.
\label{eq:residual}
\end{equation}
If correlations within a group are available, Eq.~\eqref{eq:residual} is replaced by the covariance-weighted form
\begin{equation}
r_i^{(k)}(\theta)=
\sum_{j=1}^{n_k}
\bigl[(\Sigma^{(k)})^{-1/2}\bigr]_{ij}
\bigl(y_j^{(k)}-f_j^{(k)}(\theta)\bigr),
\label{eq:residual_cov}
\end{equation}
where \(\Sigma^{(k)}\) is the covariance matrix of group \(k\) \citep{Imbrisak2025wLM}.

The corresponding group contribution to the objective is
\begin{equation}
\chi_k^2(\theta)=
\sum_{i=1}^{n_k}\bigl(r_i^{(k)}(\theta)\bigr)^2.
\label{eq:chi2k}
\end{equation}
Under the usual assumption of approximately Gaussian residuals, the negative log-likelihood is proportional, up to an additive constant, to \(\sum_{k=1}^{N_g}\chi_k^2(\theta)\) \citep{Imbrisak2025wLM}.

The Fisher Information Metric provides the metric on the statistical parameter space with respect to the parameter vector \(\theta\). In general,
\begin{equation}
g_{\mu\nu}(\theta)=
-\mathbb{E}\!\left[
\partial_\mu \partial_\nu
\log L(\theta \mid y)
\right],
\label{eq:fim_general}
\end{equation}
where \(\partial_\mu \equiv \partial/\partial\theta^\mu\) and \(L(\theta \mid y)\) is the likelihood \citep{Rao1992InformationParameters,Amari2016InformationApplications}. For nonlinear least-squares models with Gaussian residuals, the leading approximation is the Gauss--Newton form
\begin{equation}
g_{\mu\nu}(\theta)\approx
\sum_{k=1}^{N_g}\sum_{i=1}^{n_k}
\partial_\mu r_i^{(k)}(\theta)\,
\partial_\nu r_i^{(k)}(\theta).
\label{eq:fim_gn}
\end{equation}
In matrix notation this is the familiar \(J^\top J\) structure, where \(J\) is \emph{the Jacobian of residuals}. When invertible, \(g^{-1}\) gives the local covariance approximation for the parameter estimates through the Cram\'er--Rao bound \citep{Rao1992InformationParameters,Amari2016InformationApplications}.
In sloppy models, the eigenvalues of \(g_{\mu\nu}\) span many orders of magnitude, caused by strong anisotropy in parameter sensitivity and poorly constrained directions in parameter space \citep{Transtrum.2015,Imbrisak2025wLM}.

\FloatBarrier

\subsection{Weighted Fisher Information Metric}\label{sec:wFIM}

A limitation of the standard FIM is that it combines all data groups according to their raw contribution to the likelihood equally. In heterogeneous datasets, this can cause densely sampled or nominally high-precision subsets to dominate the fit. The weighted Fisher Information Metric (wFIM) addresses this by introducing a nuisance parameter \(\alpha_k\) for each group and marginalizing over these weights with a prior density \(f(\alpha_k)\) \citep{Imbrisak2025wLM}. Conditional on \(\alpha=(\alpha_1,\dots,\alpha_{N_g})\), the groupwise likelihood is
\begin{equation}
L(y \mid \theta,\alpha)=
\prod_{k=1}^{N_g}
\left(\frac{\alpha_k}{2\pi}\right)^{n_k/2}
\exp\!\left[
-\frac{1}{2}\alpha_k \chi_k^2(\theta)
\right].
\label{eq:weighted_likelihood}
\end{equation}
After marginalization over \(\alpha_k\), the effective likelihood can be written in terms of Laplace transforms,
\begin{equation}
L(\theta \mid y)=
\prod_{k=1}^{N_g}
\frac{1}{(2\pi)^{n_k/2}}
\,
\mathcal{L}_{\chi_k^2(\theta)/2}
\!\left[
\alpha_k^{n_k/2} f(\alpha_k)
\right],
\label{eq:marginal_likelihood}
\end{equation}
where the Laplace transform is
\begin{equation}
\mathcal{L}_{s}[h(\alpha)] =
\int_{0}^{\infty} e^{-s\alpha} h(\alpha)\, d\alpha.
\label{eq:laplace}
\end{equation}
If the support of \(f\) is restricted to \([\alpha_L,\alpha_U]\), the same expression applies with the integral taken over that range \citep{Imbrisak2025wLM}. This creates a straightforward way to introduce parameter constraints if necessary.

The weighted Fisher metric then takes the form
\begin{equation}
g_{\mu\nu}^{(W)}(\theta)=
\sum_{k=1}^{N_g}
w_k(\theta)\, g_{\mu\nu}^{(k)}(\theta),
\label{eq:wfim}
\end{equation}
with
\begin{equation}
g_{\mu\nu}^{(k)}(\theta)=
\sum_{i=1}^{n_k}
\partial_\mu r_i^{(k)}(\theta)\,
\partial_\nu r_i^{(k)}(\theta).
\label{eq:subset_fim}
\end{equation}
To keep the notation compact, we define
\begin{equation}
A_{m}^{(k)}(\theta)=
\mathcal{L}_{\chi_k^2(\theta)/2}
\!\left[
\alpha_k^{m} f(\alpha_k)
\right].
\label{eq:Ak}
\end{equation}
Then the general wFIM weight can be written as
\begin{equation}
w_k(\theta)=
\frac{A_{n_k/2+1}^{(k)}}{A_{n_k/2}^{(k)}}
+
\frac{A_{n_k/2+2}^{(k)}}{A_{n_k/2}^{(k)}}
-
\left(
\frac{A_{n_k/2+1}^{(k)}}{A_{n_k/2}^{(k)}}
\right)^2.
\label{eq:general_weight}
\end{equation}
This expression shows that the contribution of each group is controlled jointly by its size \(n_k\), its current value of \(\chi_k^2(\theta)\), and the chosen prior over \(\alpha_k\). The resulting geometry therefore adapts during optimization rather than remaining fixed \citep{Imbrisak2025wLM}.

We consider two prior models for \(\alpha\): a Poisson distribution and an extreme-value distribution. The Poisson prior is appropriate when \(\alpha\) is treated as a discrete count variable, which is a natural choice when radio observations lie close to the root-mean-square noise of the corresponding VLA and GMRT maps. In that case,
\begin{equation}
    f_{\mathrm{P}}(\alpha)=\frac{\mu^\alpha e^{-\mu}}{\alpha!}
\end{equation}
with \(\alpha\) a discrete non-negative integer and \(\mu>0\). As a continuous alternative, motivated by the large spread of the RxAGN data points below $1\,\mathrm{GHz}$, we also consider the extreme-value distribution (EVD),
\begin{equation}
    f_{\mathrm{EVD}}(\alpha)=\frac{1}{\sigma}\exp\!\left[-\frac{\alpha-\mu}{\sigma}-\exp\!\left(-\frac{\alpha-\mu}{\sigma}\right)\right],
\end{equation}
where \(\mu\) is the location parameter and \(\sigma>0\) is the scale parameter. The Poisson prior enters Eq.~\eqref{eq:laplace} through a discrete sum, whereas the extreme-value prior is incorporated through the corresponding continuous integral.

In Fig.~\ref{fig:prior_weights}, we compare the behavior induced by the two prior models for groups of sizes 5, 50, and 100. The upper panel corresponds to the extreme-value-distribution prior and the lower panel to the Poisson prior. Within each panel, the top row shows the negative log-likelihood for each group, and the bottom row shows the corresponding FIM weights. In both cases, the negative log-likelihood approaches the unweighted behavior at {large-$\chi_k^2$}, while the prior parameters primarily affect the small-$\chi_k^2$ regime. Likewise, the weights $w_k$ tend to unity for large $\chi_k^2$ and approach prior-dependent plateaus for small $\chi_k^2$, showing how both priors modulate the contribution of low-information groups to the effective information geometry. The detailed transition differs between the panels because the extreme-value prior is controlled by the continuous location-scale pair $(\mu,\sigma)$, whereas the Poisson prior is controlled by the discrete-count mean $\mu$.

\subsection{FBET}\label{sec:FBET}
FBET combines prior information with measurement covariance in a linearized Bayesian update \citep{leeb_consistent_2008}. In that setting, a prior mean \(y_0\) and prior covariance \(A_0\) are combined with measurements \(y_m\) and measurement covariance \(B\). After linearization with a sensitivity operator \(S\), the posterior mean and covariance are given by
\begin{align}
y_1 &=
y_0 + A_0 S^\top
\bigl(SA_0S^\top + B\bigr)^{-1}
(y_m - Sy_0),
\label{eq:fbet_mean}\\
A_1 &=
A_0 - A_0 S^\top
\bigl(SA_0S^\top + B\bigr)^{-1}
S A_0.
\label{eq:fbet_cov}
\end{align}
The corresponding $\chi^2$ value is
\begin{equation}
    \chi^2=(y_m - Sy_0)^TQ^{-1}
(y_m - Sy_0),
\end{equation}
where we used the abbreviation $Q=SA_0S^\top + B$. 
Following  \citet{schnabel_fitting_2018}, the sensitivity matrix \(S\) can be constructed by representing the quantity of interest on an arbitrary grid as a normalized Gaussian-kernel average of the measured data. If the measurements are \(y_m=(y_1,\dots,y_n)\) at points \(x_i\), then the value at a grid point \(\xi\) is written as
\[
f_M(y)=\frac{\sum_i y_i\,\phi(\xi\mid x_i,w)}{\sum_i \phi(\xi\mid x_i,w)}=S^\top y,
\]
where the kernel functions are Gaussians centered at the measurement locations,
\[
\phi(\xi\mid x_i,w)=\frac{1}{\sqrt{2\pi}w}
\exp\!\left[-\frac{(\xi-x_i)^2}{2w^2}\right].
\]
The sensitivity matrix \(S\) comprises normalized Gaussian weights that quantify how strongly each grid point is influenced by nearby measurements, while the bandwidth \(w\) sets the smoothing scale.

\subsection{Covariance Minimization Algorithm}\label{sec:cov_min}

The covariance minimization step refines the effective measurement covariance \(Q\) used in the overall \textsc{FIMER} framework described in the next section, starting from the weighted prior \(Q^{(w)}\). This update produces a stabilized covariance estimate that is then fed back into the FBET update to compute \(y_1\) and \(A_1\); the procedure therefore closes the loop between the wFIM weighting and the Bayesian smoothing described in the \textsc{FIMER} subsection.
Figure~\ref{fig:cov_min_schema} summarizes the covariance minimization loop.
Now, let \(r \in \mathbb{R}^n\) be a residual vector and let \(Q \in \mathbb{R}^{n\times n}\) be a symmetric positive-definite covariance matrix to be optimized. The algorithm seeks an updated covariance \(Q\) that reduces the $\chi^2$ value
\begin{equation}
\chi^2(Q)=r^\top Q^{-1}r.
\end{equation}
The optimization is performed over the independent upper-triangular entries of \(Q\), preserving symmetry at every step.

The update direction is obtained from the Jacobian of the z-score \(z(Q)=Q^{-1/2}r\) with respect to the independent entries of \(Q\). Writing \(J\) for this Jacobian, the algorithm takes a damped normal-equations step for the covariance parameters,
\begin{equation}
\bigl(JJ^\top+\lambda I\bigr)x = Jr,
\label{eq:normalQ}
\end{equation}
where \(\lambda>0\) is a Levenberg-type damping parameter. The flattened solution vector \(x\in\mathbb{R}^m\) is then mapped back to a symmetric matrix \(\Delta Q\) by assigning its entries to the upper triangular part and mirroring them to the lower triangular part. 
This produces a symmetric candidate search direction in covariance space.

The outer iteration is a standard Levenberg--Marquardt loop with adaptive damping. Starting from \(Q^{(0)}=Q_0\) and \(\lambda_0\), a proposal is formed as
\begin{equation}
Q_{\mathrm{prop}}
=
Q^{(0)} + \Delta Q \,\alpha,
\label{eq:Qprop}
\end{equation}
where \(\alpha\) is the geometric scaling \citep{Imbrisak2025wLM} coefficient used to accelerate the minimization process. It is computed from the previous accepted step and the current local quadratic form,
\begin{equation}
\alpha
=
\sum_{a,b,i}
(\Delta Q_{\mathrm{prev}})_{ai}\,
(\Delta Q)_{bj}\,
(J^\top J)_{ij}.
\label{eq:alpha}
\end{equation}
The proposed matrix \(Q_{\mathrm{prop}}\) is accepted if it decreases the objective function:
\begin{equation}
\chi^2(Q_{\mathrm{prop}}) < \chi^2(Q).
\label{eq:accept}
\end{equation}
If accepted, the covariance is updated,
\begin{equation}
Q_{\mathrm{prop}}\rightarrow Q^{(1)},
\end{equation}
the damping parameter is reduced,
\begin{equation}
\frac{\lambda}{\lambda_{\mathrm{down}}}\rightarrow \lambda,
\end{equation}
and the new direction \(\Delta Q\) is stored. Otherwise, the proposal is rejected, \(Q\) is left unchanged, and the damping is increased:
\begin{equation}
\lambda\,\lambda_{\mathrm{up}}\rightarrow \lambda .
\end{equation}
The widespread standard is setting \(\lambda_{\mathrm{up}}=10\) and \(\lambda_{\mathrm{down}}=2\) \citep{Lampton1997,Tanstrum2012}. We find reliable convergence in approximately 20 iterations.

The covariance minimization routine is a Levenberg--Marquardt update applied to the covariance parameters, with an accept/reject step that adapts the damping.

\subsection{FIMER Algorithm}\label{sec:fimer}
{Here, we integrate FBET smoothing \citep{leeb_consistent_2008,neudecker_impact_2013,neudecker_advanced_2014}, which was originally developed for experimental-data evaluation and the estimation of unreported measurement correlations on an evaluation grid, with wFIM weighting and covariance minimization into a unified workflow.} Because we do not have an explicit detector model, we instead use interpolation parameters, which in our case are the Gaussian parameters of FBET. {Figure~\ref{fig:fimbet_schema} provides a schematic overview of the \textsc{FIMER} procedure.}

FBET regularizes the \emph{data representation} by combining prior structure and measurement covariance, while the wFIM regularizes the \emph{parameter-space geometry} used during nonlinear optimization. FBET-style ideas are useful when one wishes to smooth {irregular measurements obtained by different detectors} or propagate covariance to a reduced representation before model fitting. The wFIM then acts on the resulting grouped residuals during parameter estimation.
\textsc{FIMER} extends FBET by injecting wFIM-derived reliability weights into the measurement covariance while keeping the Bayesian update structure intact. The workflow is as follows. 

First, we obtain a smoothed estimate \(y_0=S S^\top y\) using Gaussian kernels. A correlation hyperparameter $\rho$ introduces a degree of freedom for modeling cross-channel coupling. The kernel bandwidth (and thus the weights) can be chosen by minimizing the cost function \(\|y-S S^\top y\|_2^2\). For two channels, cross-channel coupling is introduced through a scalar hyperparameter $\rho$: for two channels, the Gaussian weights used to build \(S\) and \(S^\top\) are multiplied across channels by a hyperparameter \(\rho\), which is optimized jointly with the Gaussian widths. Concretely, with measurements grouped into data channels \(\mathbf{x}^{(1,2)}\), grid points \(\boldsymbol{\xi}^{(1,2)}\), and Gaussian weights \(\mathbf{w}^{(1,2)}\), the kernel matrix can be written as
\begin{equation}
\Phi=
\begin{pmatrix}
\phi(\boldsymbol{\xi}^{(1)}\mid \mathbf{x}^{(1)}, \mathbf{w}^{(1)}) & \rho\,\phi(\boldsymbol{\xi}^{(1)}\mid \mathbf{x}^{(2)}, \mathbf{w}^{(1)}) \\
\rho\,\phi(\boldsymbol{\xi}^{(2)}\mid \mathbf{x}^{(1)}, \mathbf{w}^{(2)}) & \phi(\boldsymbol{\xi}^{(2)}\mid \mathbf{x}^{(2)},\mathbf{w}^{(2)})
\end{pmatrix}.
\end{equation}
Therefore, for grid points \(\boldsymbol{\xi}^{(1,2)}\) the full parameter set is \(\theta=(\mathbf{w}^{(1)},\mathbf{w}^{(2)},\rho)\), with
\begin{equation}
(\mathbf{w}^{(1)},\mathbf{w}^{(2)},\rho)=\arg\min \|y-S S^\top y\|^2.
\end{equation}
This minimization is performed using SciPy's \emph{curve\_fit} for nonlinear least-squares fitting \citep{2020SciPy-NMeth}.

Second, we compute the weighted modification to the FBET $Q$ matrix 
\begin{equation}
Q^{(w)}=S A_0 S^\top + B^{(w)}.
\end{equation}
The wFIM weights on the $(\mathbf{w}^{(1)},\mathbf{w}^{(2)},\rho)$ parameter space can be {propagated into the measurement covariance matrix, $B$,} by defining a group index map $k(i)$ that assigns each measurement $i$ to its group $k$. For a diagonal measurement covariance $B_{ij}=\sigma_i^2\delta_{ij}$, the weighted covariance retains a diagonal form,
\begin{equation}
B^{(w)}_{ij}=\frac{\sigma_i^2}{w_{k(i)}(\theta)}\,\delta_{ij},
\end{equation}
so each variance $\sigma_i^2$ is divided by the weight of the group to which measurement $i$ belongs.

Third, starting from \(Q^{(w)}\), we refine the covariance using the covariance minimization update (see Section~\ref{sec:cov_min}). Finally, we use the minimized \(Q\) {to obtain the posterior} mean \(y_1\) and covariance \(A_1\) via Eqs.~\eqref{eq:fbet_mean}--\eqref{eq:fbet_cov}. {To obtain uncertainties of the measurements themselves, we \emph{apply the pushforward operator to $A_1$ }}, \begin{equation}
Q_1=S^\top A_1 S.
\end{equation}
 The off-diagonal entries of \(Q_1\) quantify propagated correlations between measurement points, while the diagonal entries give the propagated variances at each point. {We therefore estimate the measurement uncertainties as}
\begin{equation}
\sigma=\sqrt{\mathrm{diag}(S^\top A_1 S)}.
\end{equation}

\subsection{Adaptive grid search algorithm}\label{sec:ags}

{For each data group \(i\)}, the weighting procedure based on the Poisson distribution introduces a hyperparameter \(h_i\). Thus, a (sub)dataset containing \(N_g\) groups has \(N_g\) discrete hyperparameters in total, constrained by the support of the corresponding distribution. Exhaustive exploration of this grid quickly becomes combinatorially expensive, so we use an adaptive discrete search.

The search is initialized by a coarse cross centered at
\[
h^{(c)}=\left\lfloor \frac{h^{(\max)}-h^{(\min)}}{2} \right\rfloor,
\]
together with the extreme admissible values \(h^{(\min)},h^{(\max)}\), as illustrated in Fig.~\ref{fig:adaptive_cross_sketch}. We evaluate \textsc{FIMER} on these initial candidate sets and choose the one with the smallest \(\chi^2\) value as the starting point
\[
p^{(0)}=(h_1^{(0)},\dots,h_{N_g}^{(0)}).
\]

From \(p^{(0)}\), we perform a nearest-neighbor search in the discrete hyperparameter space by evaluating points that differ by \(\pm 1\) in a single coordinate, while keeping the remaining coordinates fixed and respecting the admissible support. The candidate with the smallest \(\chi^2\) value is accepted as the next iterate,
\[
p^{(1)}=(h_1^{(1)},\dots,h_{N_g}^{(1)}).
\]

\begin{figure}[h]
    \centering
    \includegraphics[width=\linewidth]{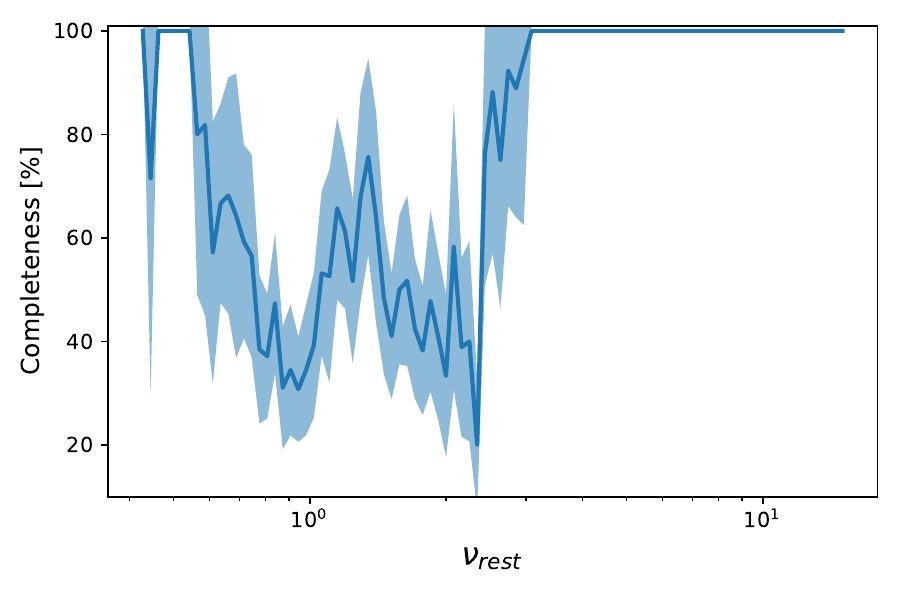}
    \caption{RxAGN dataset completeness as a function of rest-frame frequency. Shown is the percentage of detections without upper limits in the \citet{Tisanic2020TheNuclei} dataset for different rest-frame frequency bins. Where the completeness is less than 100\%, the Poissonian uncertainties for the completeness curve are shown as a shaded region.}
    \label{fig:completeness}
    
\end{figure}

\section{Data}\label{sec:data}
We evaluate \textsc{FIMER} on the RxAGN sample and flux-density measurements reported by \citet{Tisanic2020TheNuclei} for the COSMOS field. The sample is built from catalogs at four observer-frame frequencies: 325 and 610 MHz from GMRT, and 1.4 and 3 GHz from the VLA \citep{Tisanic19,Schinnerer10,Smolcic:17a}. This frequency coverage provides sub-GHz and GHz constraints for testing spectral curvature and uncertainty propagation in heterogeneous radio data.

For the VLA component, the parent RxAGN sample uses the 3 GHz VLA-COSMOS Large Project catalog together with the joint 1.4 GHz VLA-COSMOS catalog \citep{Smolcic:17a,Schinnerer10}. The 3 GHz map was built from S-band observations and source extraction was performed with \textsc{blobcat} at a 5$\sigma$ threshold \citep{Novak15,Hales12,Smolcic:17a}. The 1.4 GHz catalog is based on the VLA-COSMOS Large and Deep surveys, with source detection and measurement based on \textsc{SExtractor} and the AIPS task SAD \citep{Schinnerer07,Schinnerer10,BertinArnouts96}.

For the GMRT component, we use the 325 and 610 MHz COSMOS catalogs reduced with the SPAM pipeline, with primary-beam (and, at 610 MHz, average pointing-error) corrections applied before final catalog construction \citep{Tisanic19}. Sources were extracted with \textsc{blobcat} down to 5$\sigma$, and resolved sources were identified using the total-to-peak flux-density criterion described in \citet{Tisanic19}.

In the present work, flux densities are taken directly from these published catalogs for matched counterparts across bands (matched in \textsc{topcat}, following the RxAGN data assembly in \citet{Tisanic2020TheNuclei}). {We use the published catalog total flux densities and associated uncertainties as the input measurements for \textsc{FIMER}.}

For \textsc{FIMER}, each observing band is treated as a structured data group with its own effective uncertainty scale and possible cross-group dependence. This grouping naturally captures heterogeneity introduced by different instruments, synthesized beams, sensitivity levels, and detection completeness. 

{We represent the radio flux densities in logarithmic form to linearize the local power-law behavior of a simple single-spectral-index model \citep{Condon92}; this representation has also proved useful for fitting more complex SED parameterizations \citep{Tisanic19,Tisanic2020TheNuclei}. It improves comparability across bands and supports uncertainty propagation across the GMRT and VLA rest-frame frequency ranges.}

In Fig.~\ref{fig:completeness} we show the percentage of detections without upper limits in the \citet{Tisanic2020TheNuclei} dataset for different rest-frame frequency bins. Where completeness is significantly less than 100\%, the information derived without upper limits might be significantly skewed. We leave the explicit treatment of upper limits to future work and use the present dataset as a proof-of-principle demonstration.
\begin{figure*}[t]
\centering
    \includegraphics[width=0.9\textwidth]{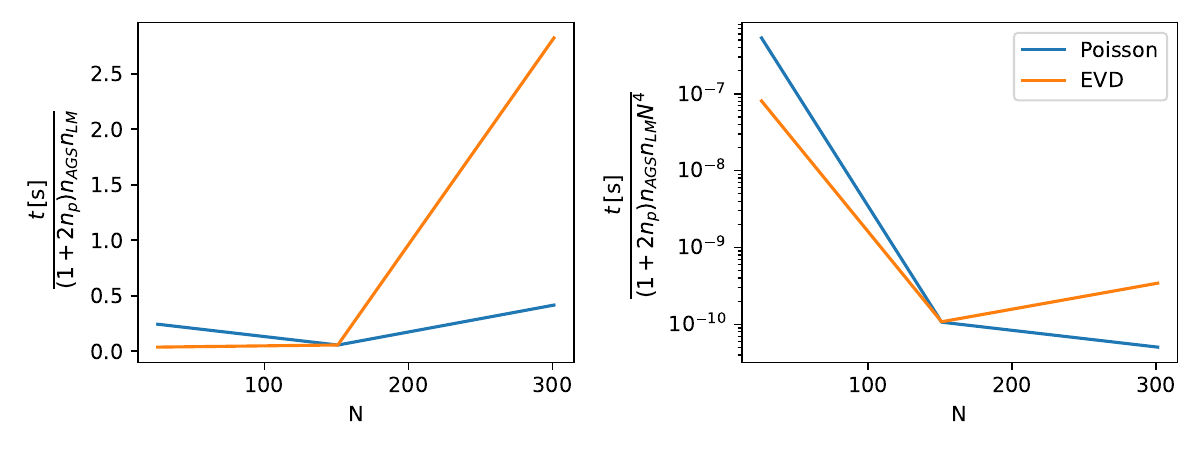}
 \caption{Runtime for the Poisson and EVD priors for different numbers of measurements, $N$. The left panel shows runtimes normalized by the estimated non-$N$ prefactor, while the right panel shows runtimes normalized by the full asymptotic scaling law, including the adaptive-search factor and the explicit $N$-dependence.}
    \label{fig:bigO}
    \centering
    \begin{subfigure}[t]{\textwidth}
        \centering
        \includegraphics[width=\linewidth]{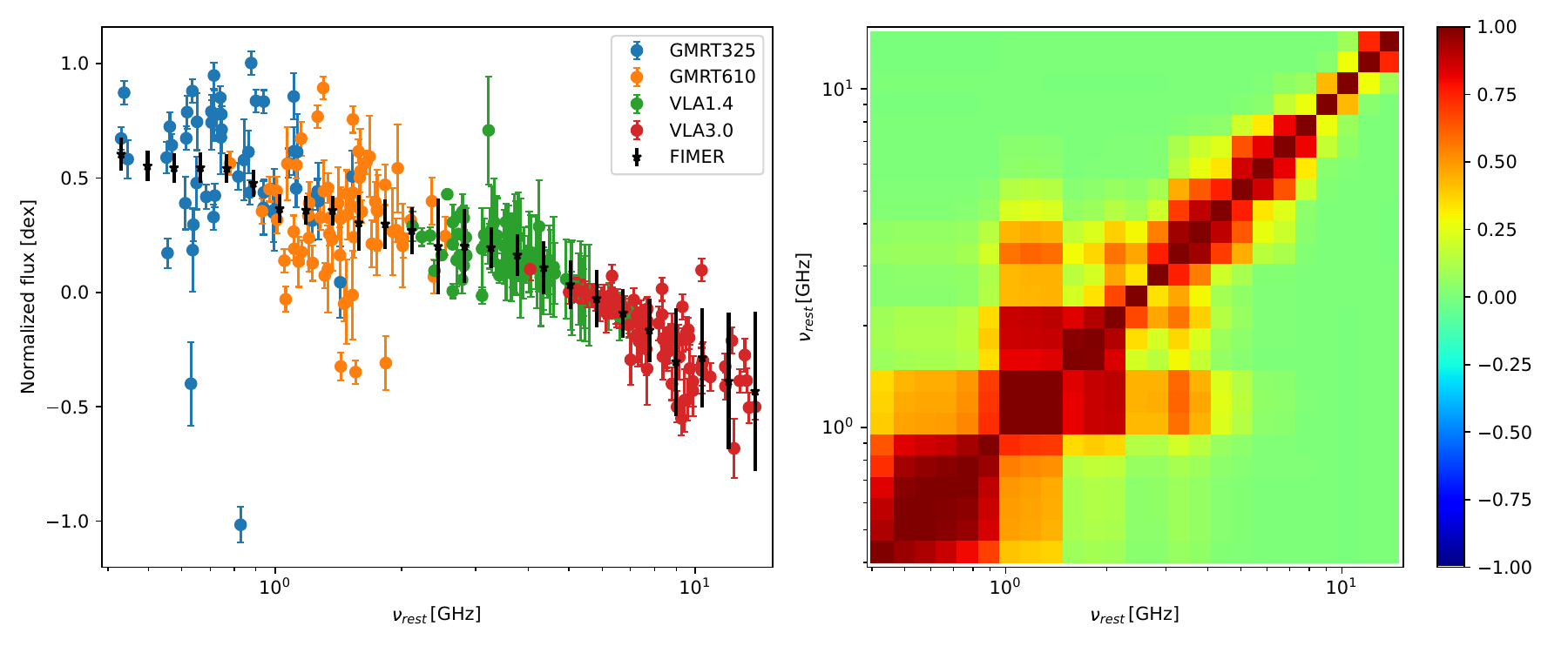}
    \end{subfigure}

    \vspace{0.5em}

    \begin{subfigure}[t]{\textwidth}
        \centering
        \includegraphics[width=\linewidth]{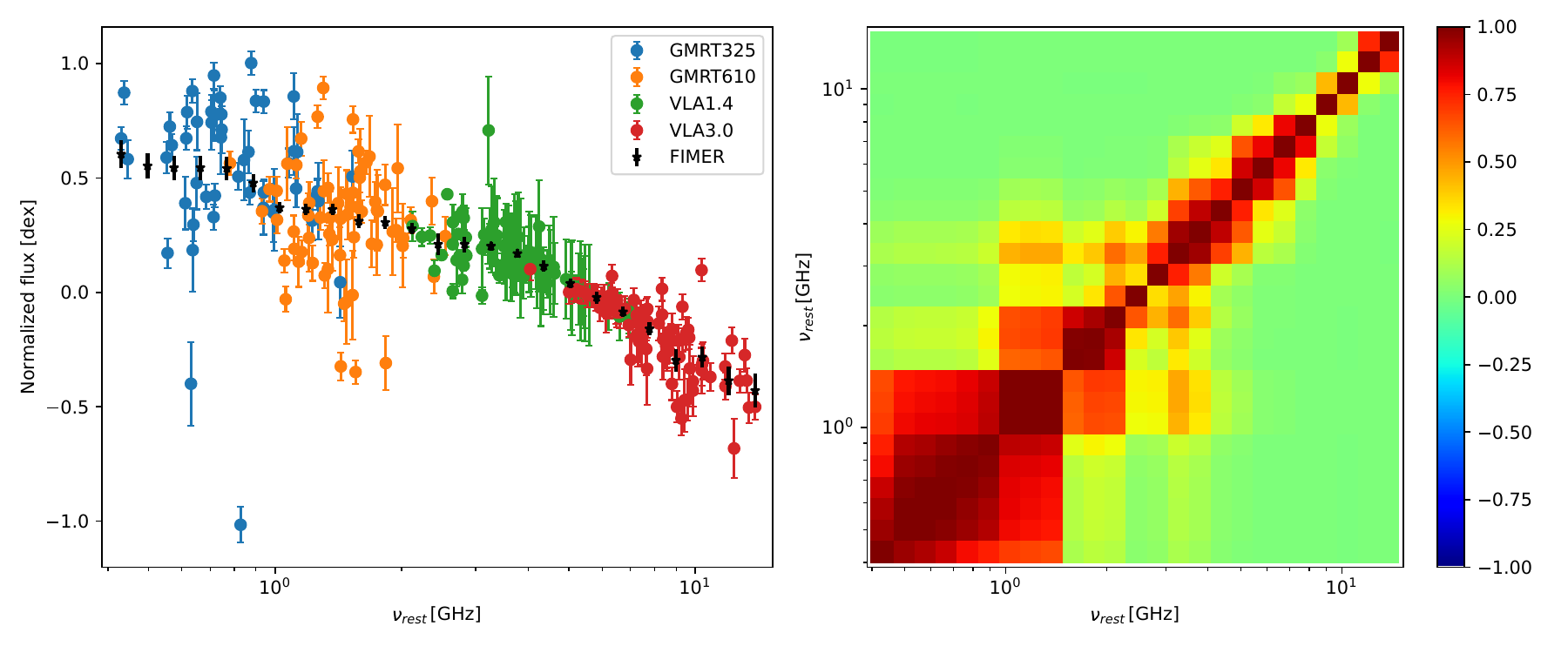}
    \end{subfigure}
    \caption{RxAGN results obtained with the extreme-value-distribution (top) and Poisson (bottom) weighting schemes. In each panel, the left-hand side shows the RxAGN radio SED for the GMRT and VLA datasets (colored points) together with the \textsc{FIMER} evaluations (black points), while the right-hand side shows the corresponding \textsc{FIMER} correlation matrix across rest-frame frequencies.}
    \label{fig:SEDs}
\end{figure*}
\begin{figure*}[t]
    \centering
    \begin{subfigure}[t]{0.49\textwidth}
        \centering
        \includegraphics[width=\linewidth]{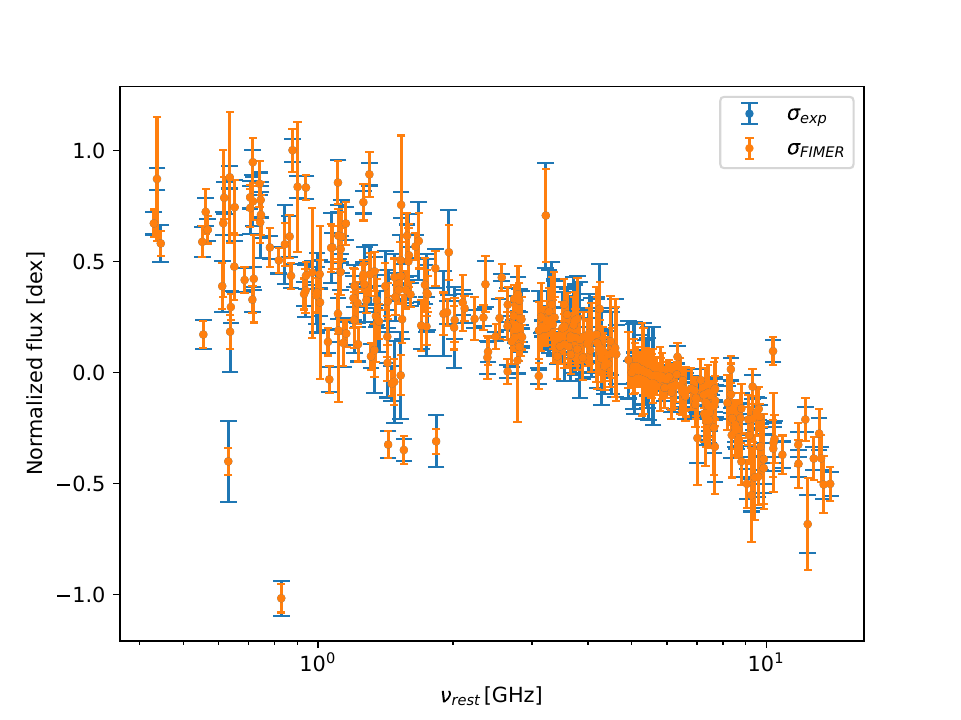}
    \end{subfigure}\hfill
    \begin{subfigure}[t]{0.49\textwidth}
        \centering
        \includegraphics[width=\linewidth]{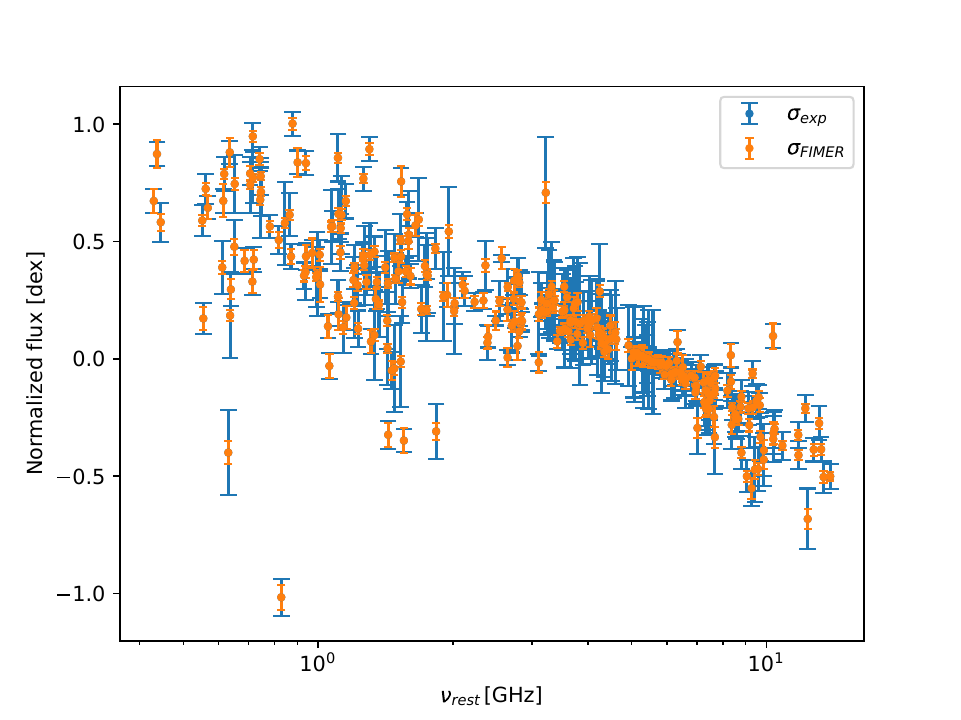}
    \end{subfigure}
    \caption*{Top row: reconstructed-versus-experimental uncertainty comparison for the extreme-value-distribution (left) and Poisson (right) weighting schemes.}

    \vspace{0.5em}

    \begin{subfigure}[t]{\textwidth}
        \centering
        \includegraphics[width=.8\linewidth,trim={0 0 14cm 0},clip]{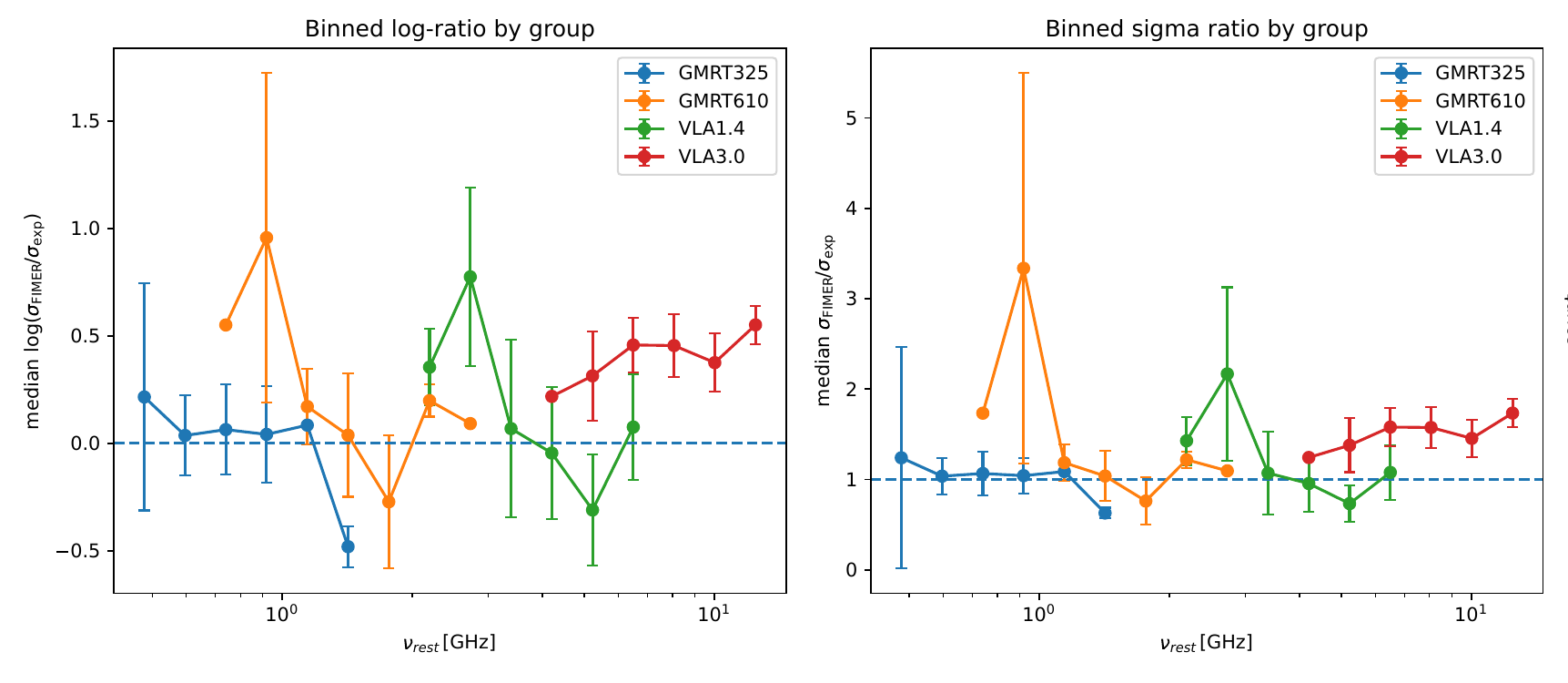}
    \end{subfigure}

    \vspace{0.5em}

    \begin{subfigure}[t]{\textwidth}
        \centering
        \includegraphics[width=.8\linewidth,trim={0 0 14cm 0},clip]{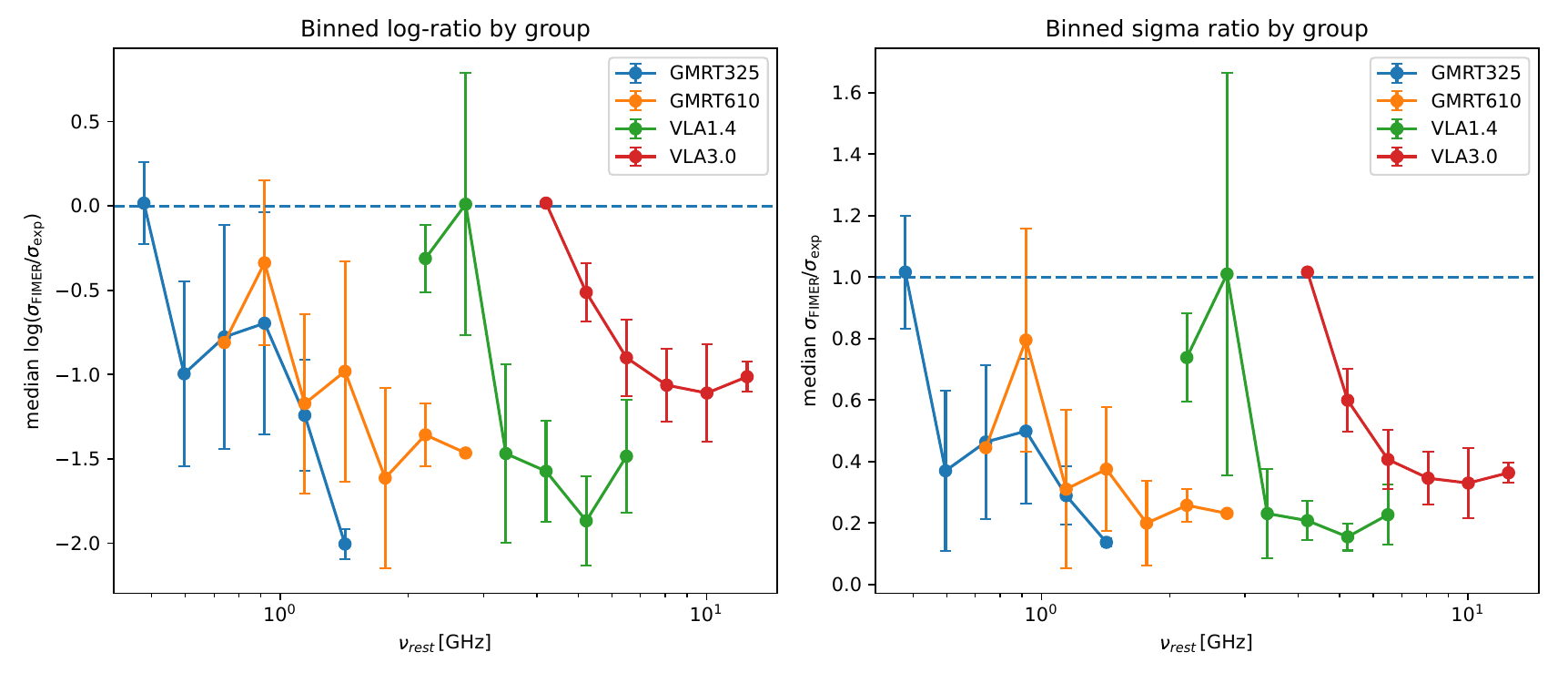}
    \end{subfigure}
    \caption*{Lower two rows: binned uncertainty diagnostics for the extreme-value-distribution (middle row) and Poisson (bottom row) weighting schemes. The left panels show the median log-ratio of \textsc{FIMER}-derived and experimental uncertainties, while the right panels show the median sigma ratio.}
    \caption{Comparison of reconstructed measurement uncertainties for the extreme-value and Poisson weighting schemes.}
    \label{fig:error_comparison}
\end{figure*}
\FloatBarrier

\section{Results and discussion}\label{sec:results}
{For visual clarity, we show results for a 20\% subsample of the original RxAGN dataset to illustrate the feasibility of \textsc{FIMER}. Figure~\ref{fig:bigO} shows the runtimes for the Poisson and EVD priors for different measurement-subset sizes, $N$. We estimate the leading-order computational complexity as $O\left((1+2n_p)n_{AGS}n_{LM}N^4\right)$, where $n_p=1+n_G n_f$ is the number of prior parameters, namely the Tikhonov regularization parameter $\lambda$ and $n_f$ prior-distribution parameters for each of the $N_g$ data groups. Here, $n_{AGS}$ denotes the number of adaptive grid-search steps, taken from the measured runtimes in Fig.~\ref{fig:bigO}, and $n_{LM}$ denotes the preset maximum number of Levenberg--Marquardt steps in the covariance minimization. This scaling law should be regarded as a rough estimate, because the algorithm can converge before reaching the maximum number of steps. The left panel is included because, for smaller $N$, the $N^4$ factor is not yet dominant, which leads to a flatter dependence of the Poisson-prior runtime on $N$. This runtime model also does not account for the effects of sparse-matrix linear algebra, which we use to avoid memory bottlenecks for large tensors.} 

In Fig.~\ref{fig:SEDs}, we show the RxAGN results obtained with the extreme-value and Poisson weighting schemes. In this demonstration, \textsc{FIMER} is initialized without prior knowledge of individual measurement uncertainties, which are set to unity in the FBET stage. The panels show the \textsc{FIMER} evaluations together with the correlation coefficients evaluated {on the rest-frame frequency grid}. The binning strategy of \citet{Tisanic2020TheNuclei} is useful for capturing the overall behavior when upper limits are included, but it cannot recover correlations between rest-frame frequency bins. As shown in Fig.~\ref{fig:SEDs}, non-adjacent rest-frame frequencies are correlated, particularly below $\sim 3\,\mathrm{GHz}$. Above $\sim 3\,\mathrm{GHz}$, correlations are less confined to neighboring bins, where the VLA 3 GHz dataset begins to dominate, likely owing to overlap among the observer-frame frequency datasets. Although often neglected, such correlations can produce unexpected means and variances, known as the Peelle's Pertinent Puzzle, an effect known to be important in other fields such as nuclear data evaluation \citep{Neudecker2014}.

Figure~\ref{fig:error_comparison} presents the main result of this demonstration: reconstruction of effective measurement uncertainties and their comparison with catalog-reported uncertainties. For the Poisson distribution, the reconstructed measurement uncertainties are systematically underestimated, particularly for sources at higher redshifts, over all GMRT and VLA datasets. The presence of a large spread of normalized fluxes at lower rest-frame frequencies motivated us to consider the extreme-value distribution. We find that this prior distribution reconstructs the measurement uncertainties well over the entire rest-frame frequency range. Since the extreme-value distribution is related to the Weibull distribution, used in \citet{Tisanic2020TheNuclei} to describe this RxAGN sample in the presence of upper limits, this would be a useful further line of work.

\FloatBarrier

\section{Conclusion}\label{sec:conclusion}
We presented \textsc{FIMER}, an information-geometric framework that combines wFIM weighting and FBET updates to {reconstruct effective measurement uncertainties and correlations} for heterogeneous astrophysical data. In the RxAGN demonstration, the method recovered stable uncertainty patterns across frequency bins. These results highlight the importance of explicitly modeling heterogeneity and covariance when detector-level uncertainty information is incomplete.

{A central feature of \textsc{FIMER} is that the weighting priors are not arbitrary regularization choices. Instead, they provide a way to incorporate prior statistical knowledge about detector behavior directly into the reconstruction procedure. In this sense, the method is statistically interpretable: Poisson-type priors encode counting-statistics intuition, while extreme-value priors allow tail-dominated or asymmetric fluctuations to be represented when such effects are expected to influence the inferred distribution of measurement uncertainties. The resulting reconstruction is therefore guided by explicit and testable assumptions about the measurement process rather than by purely empirical tuning.}

\textsc{FIMER} {thus offers a principled technique for reconstructing measurement uncertainties when they are unavailable or are known to be underestimated or correlated. More broadly, the method shows that the reconstruction of measurement uncertainties can itself be treated as a statistical inference problem in heterogeneous astrophysical datasets, provided that physically motivated assumptions about detector statistics are made explicit.}

A natural next step is to extend the framework to incorporate upper limits directly and to validate it on larger samples. {Further developments should also explore broader prior families, richer cross-group covariance parameterizations, and applications to other multi-instrument datasets in which incomplete uncertainty information limits statistical analysis.}
\section*{Acknowledgements}
The authors are grateful to Zrinka Vidovi\'{c} Tisani\'{c} for additional comments and insightful discussions that helped improve this manuscript.

\bibliographystyle{aa}
\bibliography{references}

\end{document}